 
\input harvmac
\input epsf

\overfullrule=0pt
 
\def\A{{\scriptscriptstyle A}}

\def\R{{\scriptscriptstyle R}}

\def\T{{\scriptscriptstyle T}}


\def\CN{{\cal N}}

 
\def\a{\alpha}
\def\b{\beta}

\def\e{\epsilon}


\def\half{{1 \over 2}}
\def\quarter{{1 \over 4}}

\def\third{{1 \over 3}}
\def\threehalves{{3 \over 2}}


\def\bar#1{\overline{#1}}
\def\BoldC{\bf C}
\def\ccdot{\hbox{\kern-.1em$\cdot$\kern-.1em}}

\def\Gdual{{\widetilde G}}
\def\gtap{\raise.3ex\hbox{$>$\kern-.75em\lower1ex\hbox{$\sim$}}}

\def\ltap{\raise.3ex\hbox{$<$\kern-.75em\lower1ex\hbox{$\sim$}}}
\def\Nc{{N_c}}
\def\Nf{{N_f}}
\def\R{{\scriptscriptstyle R}}

\def\SpN{Sp(2N)}
\def\SON{SO(N)}

\def\ev#1{\langle#1\rangle}
\def\therefore{{\hbox{..}\kern-.43em \raise.5ex \hbox{.}}\>\>}
\def\Wtree{W_{\rm tree}}
\def\Wdualtree{{\widetilde W}_{\rm tree}}
\def\Wdyn{W_{\rm dyn}}
\def\Wtot{W_{\rm tot}}


\def\sym{  \> {\vcenter  {\vbox  
              {\hrule height.6pt
               \hbox {\vrule width.6pt  height5pt  
                      \kern5pt 
                      \vrule width.6pt  height5pt 
                      \kern5pt
                      \vrule width.6pt height5pt}
               \hrule height.6pt}
                         }
              } \>
           }
\def\fund{  \> {\vcenter  {\vbox
              {\hrule height.6pt
               \hbox {\vrule width.6pt  height5pt
                      \kern5pt
                      \vrule width.6pt  height5pt }
               \hrule height.6pt}
                         }
                   } \>
           }
\def\anti{ \>  {\vcenter  {\vbox
              {\hrule height.6pt
               \hbox {\vrule width.6pt  height5pt
                      \kern5pt
                      \vrule width.6pt  height5pt }
               \hrule height.6pt
               \hbox {\vrule width.6pt  height5pt
                      \kern5pt
                      \vrule width.6pt  height5pt }
               \hrule height.6pt}
                         }
              } \>
           }

\newdimen\pmboffset
\pmboffset 0.022em
\def\oldpmb#1{\setbox0=\hbox{#1}%
 \copy0\kern-\wd0 \kern\pmboffset\raise
 1.732\pmboffset\copy0\kern-\wd0 \kern\pmboffset\box0}

\nref\reviews{For reviews, see K. Intriligator and N. Seiberg,
hep-th/9509066, Nucl. Phys.  Proc. Suppl. {\bf 45BC} (1996)~1 ;
 M.E. Peskin, hep-th/9702094 ; M. Shifman, hep-th/9704114,
 Prog. Part. Nucl. Phys. {\bf 39} (1997) 1.}
\nref\SUSYbreakreviews{For 
recent reviews of dynamical supersymmetry 
  breaking, see W. Skiba,  
hep-th/9703159; A. Nelson, hep-ph 9707442;
  E. Poppitz, hep-ph/9710274;
  S. Thomas, hep-th/9801007.}
\nref\ISS{K. Intriligator, 
N. Seiberg and S. Shenker, hep-th/9410203, 
Phys. Lett. {\bf 
   B342} (1995) 152.}
\nref\Manohar{G. Dotti and A.V. Manohar, hep-th/9712010.}
\nref\CSST{C. Csaki, M. Schmaltz, W. Skiba, and J. Terning,
  hep-th/9801207.}
\nref\Preskill{J. Preskill, S. Trivedi, 
F. Wilczek and M. Wise, Nucl. Phys.  {\bf B363} (1991) 207.}
\nref\Ibanez{L. Ib\'a\~nez and G. Ross, Phys. Lett. 
{\bf 260B} (1991) 291;
  Nucl. Phys. {\bf B368} (1992) 3; L. Ib\'a\~nez, 
Nucl. Phys. {\bf B398}
  (1993) 301.}
\nref\Banks{T. Banks and M. Dine, Phys. Rev. {\bf D45} (1992) 1424.}
\nref\CsakiMurayama{C. Cs\`aki and H. Murayama, hep-th 9710105.}
\nref\tHooft{G.~'t Hooft, Phys. Rev. Lett. {\bf 37} (1976) 8; 
Phys. Rev. {\bf
  D14} (1976) 3432.}
\nref\ISthree{K. Intriligator and N. Seiberg, hep-th/9503179, 
Nucl. Phys. {\bf B444} 
(1995) 125.}
\nref\IStwo{K. Intriligator and N. Seiberg, 
hep-th/9506084, {\it Proc. of
Strings '95}, edited by I. Bars et. al., (World Scientific, 1996)}
\nref\Intriligator{K. Intriligator, hep-th/9505051,
Nucl. Phys. {\bf B448} (1995) 187.}
\nref\Leigh{R.G. Leigh and M.J. Strassler, hep-th/9503121,
Nucl. Phys. {\bf B447} (1995)
  95.}
\nref\Parisi{G. Parisi and N. Sourlas, Phys. Rev. Lett. {\bf 43} 
(1979) 744.}
\nref\Penrose{R. Penrose, in {\it Combinatorial Mathematics and its 
  Applications}, edited by  D.J.A Welsh (Academic Press, New York, 
1971).}
\nref\Cvitanovic{P. Cvitanovic, {\it Group Theory}, Nordita Notes 
(1984) 136.}
\nref\King{R.C. King, Can. J. Math. {\bf 23} (1971) 176.}
\nref\Kennedy{P. Cvitanovic and 
A.D. Kennedy, Phys. Scr. {\bf 26} (1982) 
  5.}
\nref\Dunne{G.V. Dunne, J. Phys. A: 
Math. Gen. {\bf 22} (1989) 1719.}
\nref\ChoKraus{P. Cho and P. Kraus, hep-th/9607200,
Phys. Rev. {\bf D54} (1996) 7640.}
\nref\CSS{C. Cs\`aki, M. Schmaltz and W. Skiba, hep-th/9607210, 
Nucl. Phys.  {\bf B487} (1997) 128.}

\def\LongTitle#1#2#3#4#5{\nopagenumbers\abstractfont
\hsize=\hstitle\rightline{#1}
\hsize=\hstitle\rightline{#2}
\hsize=\hstitle\rightline{#3}
\vskip 0.5in\centerline{\titlefont #4} \centerline{\titlefont #5}
\abstractfont\vskip .3in\pageno=0}

\LongTitle{HUTP-98/A001}{PUPT-1758}{UCSD/PTH 98-04}
{Misleading Anomaly Matchings?}{}

\centerline{
  John Brodie${}^1$, Peter Cho${}^2$ and Kenneth Intriligator${}^3$}
\bigskip
\centerline{${}^1$ Department of Physics, Princeton University,
  Princeton, NJ 08540}
\centerline{${}^2$ Lyman Laboratory, Harvard University, 
Cambridge, MA  
  02138}
\centerline{${}^3$ Department of Physics, University of 
California at San 
  Diego,}
\centerline{9500 Gilman Drive, La Jolla, CA  92093}

\vskip 0.3in
\centerline{\bf Abstract}
\bigskip

We investigate the low energy dynamics of $\CN=1$ supersymmetric
$\SON$ gauge theories with a single symmetric tensor matter field.
These theories exhibit non-trivial matching of global 't Hooft
anomalies at the origin of moduli space.  We argue that their quantum
moduli spaces possess distinct Higgs and confining branches which
touch at the origin in an interacting non-Abelian Coulomb phase.  The
matching of anomalies between microscopic degrees of freedom and
colorless moduli therefore appears to be coincidental.  We discuss a
formal mathematical relation between the $\SON$ model and an analogous
$\SpN$ theory with a single antisymmetric matter field which provides
an explanation for the anomaly matching coincidence.

\Date{2/98}

	Recent advances in analyzing the strong coupling dynamics of
supersymmetric gauge theories have opened up several new directions
for model building \reviews.  One interesting application has been to
the study of dynamical supersymmetry breaking.  A number of theories
utilizing novel dynamical mechanisms to break SUSY have been
constructed during the past few years \SUSYbreakreviews.  One
especially simple proposal is based upon an $SU(2)$ model with a
single isospin-3/2 matter field
\ISS.  Nontrivial `t~Hooft anomaly matching in this chiral model
suggests that it confines with a smooth quantum moduli space.
Supersymmetry breaking then results upon adding a tree level
superpotential.  However, there is another possibility for the $SU(2)$
model's low energy dynamics \ISS: the 't Hooft anomaly matching may be
coincidental and the theory may have a non-trivial RG fixed point at
the origin.  In this case, supersymmetry need not be broken upon
adding the tree level superpotential.  Which scenario is correct
remains an unsettled question.

	In this note, we examine a simple class of $\CN=1$
supersymmetric $SO(N)$ models with a single matter chiral superfield
$S$ in the two-index, symmetric, traceless, tensor representation
$\sym$.  As we shall see, these models are similar to the $SU(2)$
theory inasmuch as they exhibit non-trivial 't Hooft anomaly matching,
which suggests confinement with a smooth moduli space.
\foot{ These models were recently considered in \Manohar\ as part
of a complete classification of all theories based on simple gauge
groups with a freely generated moduli space and matching 't~Hooft
anomalies.  Some of our previously unpublished observations on these
models were cited in that work.  The $\SON$ theories were also
recently constructed via branes in \CSST.}
We will argue, however, that the moduli space for these theories {\it
must} have a more intricate structure, with various branches and a
non-trivial RG fixed point at the origin.  The anomaly matching then
appears to be coincidental.  A skeptic might view the $\SON$ models as
casting doubt on the proposed confinement and supersymmetry breaking
of the model of \ISS.  At the very least, they demonstrate that
anomaly matching can be misleading.

	It is important to recall that the $S$ field does not
transform according to a faithful representation of the $\SON$ gauge
group's center.  Test charges in spinor or vector representations
cannot be screened by either gluons or dynamical $S$ matter fields.
Therefore, the $\SON$ model's moduli space can {\it \`a priori} have
distinct Higgs, confining and oblique confining branches where Wilson
and `t~Hooft loops exhibit various possible area and perimeter law
scalings.  This feature of the $\SON$ model represents a clear
qualitative difference with the $SU(2)$ theory of \ISS\ whose $I=3/2$
field does provide a faithful representation of the $Z_2$ center of
$SU(2)$.

	In the absence of any tree level superpotential, the $\SON$
model has an anomaly free $U(1)_\R$ symmetry with $R(S)=4/(N+2)$.
\foot{We adopt the $\SON$ index values $\mu (\fund) = 2$, 
$\mu({\rm Adj}=\anti) 
= 2N-4$ and $\mu (\sym)~=~2N+4$ which count numbers of fermion zero
modes in a single instanton background.}
Its one loop beta function is $b_0=2(N -4)$, so the $N \ge 5$ theories
are asymptotically free.  Although $b_0=0$ for the $SO(4)\cong SU(2)
\times SU(2)$ model with $S \sim ({\bf 3,3})$, this theory is not
asymptotically free at two loop order.  It thus flows to a free theory
in the IR.  Similarly, the $SO(2)$ and $SO(3)$ theories are not
asymptotically free and flow to free theories in the IR.

	The $\SON$ model possesses a moduli space of classical vacua
given by solutions to the $D$-flatness condition modulo gauge
transformations.  $D_a=\Tr(T_a S S^\dagger)=0$ implies that the real
and imaginary parts of $S$ commute and can be simultaneously
diagonalized by an $\SON$ rotation.  The moduli space is consequently
$N-1$ complex dimensional.  Throughout its bulk, the gauge group is
generically completely broken by the Higgs mechanism.  But there exist
subspaces of enhanced gauge symmetry where some diagonal expectation
values of $S$ are equal.  On the subspace where
\eqn\generalSvev{\langle S \rangle = 
\pmatrix{z_1 {\bf 1}_{m_1 \times m_1} &      			 
&        & \cr
                                & z_2 {\bf 1}_{m_2 \times m_2} & & \cr
 			        & & \ddots & \cr & & & z_\ell {\bf
 			        1}_{m_\ell \times m_\ell} \cr}}
with $\sum_{i=1}^\ell m_i = N$ and $\sum_{i=1}^\ell m_i z_i = 0$, the
low energy theory reduces to $l$ decoupled $SO(m_i)$ models, each with
a traceless two index symmetric tensor $\sym _i$, and $l-1$ singlet
moduli corresponding to the $z_i$.

	The $N-1$ dimensional classical moduli space is freely
generated by arbitrary expectation values of the gauge invariant
operators
\eqn\Oops{O_n = \Tr S^n, \qquad n=2,3,\cdots, N.}
One can also form the additional composite $B = \det S$.  But it is
{\it linearly} related by the trace of the characteristic polynomial for 
matrix $S$
\eqn\charpoly{O_{N} - \half O_2 O_{N-2} - \third O_3 O_{N-3} + 
\cdots + (-1)^{N} N B = 0}
to the operators in \Oops.  This last expression can be used to
eliminate $B$ regardless of any possible quantum corrections.

	In the quantum theory, any dynamically generated
superpotential which could lift the classical moduli space degeneracy
is determined by holomorphy and symmetry considerations to be of the
form
\eqn\Wdynamical{\Wdyn = C \Bigl[{S^{2N+4} \over \Lambda^{2N-8}} 
\Bigr]^\quarter}
where $\Lambda$ denotes the $\SON$ scale and $S^{2N+4}$ stands for
some function of the $O_n$ operators which has $S$ number equal to
$2N+4$.  Asymptotic freedom requires the classical moduli space to be
recovered in the weak coupling $\ev{S}/\Lambda \rightarrow \infty$
limit.  This condition is incompatible with the form of $\Wdyn$ which
yields a potential that grows with $S$.  Consequently, the constant
$C$ must vanish.  We will refer to this part of the moduli space where
$\Wtot=\Wdyn+\Wtree=0$ as the ``Higgs branch.''  Shortly, we will
argue that the $\Wdyn$ in \Wdynamical\ with $C\neq 0$ must be
generated on another ``confining branch'' of the theory when $\Wtree
\neq 0$.

	Although strong dynamics do not generate any superpotential on
the Higgs branch, they can still alter the theory's vacuum structure
and lead to interesting phenomena near the origin.  Classically, the
moduli space metric for the $O_n$ fields has singularities on
subspaces of the vacuum manifold where the gauge group is not
completely broken.  In the classical theory, massless $\sym _i$ matter
fields and $SO(m_i)$ gluons must be included in order to obtain a
non-singular description.  In the quantum theory, the moduli space
singularities are either smoothed out or else reflect possibly
different massless fields.

	Since the global $U(1)_\R$ symmetry remains unbroken at the
origin, it is possible to check 't Hooft anomalies in order to
constrain the massless spectrum at this point.  In the microscopic
theory, the $U(1)_\R$ and $U(1)^3_\R$ anomalies assume the values
\eqn\partonanomalies{\eqalign{
A_{U(1)_\R} &= \Bigl[ {N+1 \choose 2} -1 \Bigr] 
 \Bigl[-{N-2 \over N+2} \Bigr] + {N \choose 2}[1] = N-1 \cr
A_{U(1)^3_\R} &= \Bigl[ {N+1 \choose2}-1 \Bigr]
 \Bigl[-{N-2 \over N+2} \Bigr]^3 + {N \choose 2}[1]^3 = 
 {(N-1)(5 N^2 - 4 N+ 4) \over (N+2)^2}. \cr}}
In the effective theory, we find the following contributions from the
fermionic components of the $O_n$ moduli superfields:
\eqn\hadronanomalies{\eqalign{
A_{U(1)_\R} &= \sum_{n=2}^{N} 
\Bigl[ {4 n \over N+2}-1 \Bigr] = N-1 \cr
A_{U(1)^3_\R} &= \sum_{n=2}^{N} 
\Bigl[ {4 n \over N+2}-1 \Bigr]^3 = 
 {(N-1)(5 N^2 - 4 N+ 4) \over (N+2)^2}. \cr}}
The anomalies precisely match!  This nontrivial agreement suggests
that the $O_n$ composites saturate the massless spectrum of the
quantum theory.  If so, the quantum Kahler metric for the $O_n$ moduli
should be flat near the origin and non-singular throughout the moduli
space.  This anomaly matching represents circumstantial evidence for
confinement in the $\SON$ model in the same way as for the $SU(2)$
model of \ISS.

It is interesting to further consider discrete anomalies 
\refs{\Preskill{--}\Banks}.  Global anomalies for 
$Z_N$ groups should match between high and low energy descriptions of
any gauge theory
\CsakiMurayama.  In particular, the $Z_N$-gravity-gravity coefficients
are supposed to agree modulo $N$ (modulo $N/2$) for $N$ odd (even).
On the other hand, other anomalies such as $Z_N^3$ and $Z_N^2 U(1)_R$
can be corrupted by unknown massive state contributions.  Therefore,
they cannot be used to definitively rule out a proposed massless
confining phase spectrum.

        In the $\SON$ model, instantons break the classical $S$-number
symmetry down to a $Z_{2N+4}$ subgroup \tHooft.
\foot{A $Z_{(N+2)/4}$ subgroup of $Z_{2N+4}$ is contained within 
$U(1)_\R$.  But there is no loss in working with the larger $Z_{2N+4}$
symmetry since $U(1)_\R$ anomalies are already known to match.}
After assigning the $S$ field charge $1$ under this discrete group, we
find that the $Z_{2N+4}^3$, $Z_{2N+4}^2 U(1)_\R$ and $Z_{2N+4}
U(1)_\R^2$ anomalies do not match for general $N$.  However, the
$Z_{2N+4}$ anomalies calculated in the microscopic gauge theory and
low energy sigma model are equal:
\eqn\Zgravgravanom{A_{Z_{2N+4}} = \Bigl[ {N(N+1) \over 2} -1 \Bigr] (1)
= \sum_{n=2}^N n = {(N-1)(N-2) \over 2}.}
These discrete anomaly matching results are therefore consistent with
the hypothesis that the $\SON$ model confines and yields a free field theory 
at the origin.

	Although the picture of simple confinement at the origin
passes all non-trivial anomaly tests, it can {\it not} provide a
complete description of the moduli space.  The first problem arises
along subspaces of enhanced gauge symmetry within the Higgs branch
where $\SON \to \prod_i SO(m_i)$ with some $m_i=2$, 3 or 4.  On these
subspaces, the low energy theory is not asymptotically free.  The
massless spectrum must therefore include $SO(m_i)$ gauge fields and
$\sym _i$ matter fields.  Since these subspaces intersect the origin,
the massless spectrum at $\vev{S}=0$ can not simply consist of the
confined $O_n$ moduli.  We note that no such free-electric subspaces
enter into the $SU(2)$ model of \ISS\ or other theories which are
believed to confine.

	Another way in which there could exist a free theory at the
origin would be to have the same massless spectrum at $\vev{S}=0$ as
that which exists on the free-electric subspace with maximal unbroken
gauge group.  For example when $N$ is a multiple of four, the massless
spectrum on the free-electric subspace with maximal unbroken gauge
group consists of free-electric $SO(4)^{N/4}$ gauge fields and
symmetric tensors, along with a subset of the $O_n$ operators which
parameterize this subspace.  This massless spectrum could conceivably
extend down to the origin of moduli space.  However, there seems to be
no way to make this scenario compatible with the required 't~Hooft
anomaly matching at the origin.  It thus appears that the theory at
the origin can not be free and that the 't~Hooft matching observed
above is simply a misleading coincidence.

	Another reason why confined $O_n$ fields cannot represent the
complete massless spectrum at the origin of moduli space may be seen
by turning on a tree level superpotential
\eqn\addw{\Wtree =\sum _{n=2}^{N} g_n O_n.}
In the presence of these classical terms, the eigenvalues $z$ of $S$ 
represent different solutions of $\sum_{n=1}^N n g_n z^{n-1}=0$. 
\foot{The coupling $g_1$ is a Lagrange multiplier which enforces the 
tracelessness condition for $S$.}
Various vacua exist where different numbers of eigenvalues are equal.
The general low energy theory is $\CN=1$ pure Yang-Mills with a
$\prod_i SO(m_i)$ gauge group.  For $m_i \ge 3$, $SO(m_i)$ confines
and yields $m_i-2$ supersymmetric vacua when $m_i \ge 5$, four vacua
when $m_i=4$ and two vacua when $m_i=3$.  The low energy spectrum
contains a photon for each $m_i=2$.

For simplicity, consider the special case where only the mass coupling
$g_2 = \half m$ in \addw\ is nonzero.  For $m \gg
\Lambda$, the heavy $S$ can be integrated out.  The resulting low
energy $SO(N)$ Yang-Mills theory is known to confine with a mass gap
and to have $N-2$ supersymmetric vacua.  If the original $\SON$ model
simply had a moduli space with confinement in terms of the $O_n$
fields, turning on $\Wtree=\half m O_2$ would lead to dynamical SUSY
breaking in a similar fashion to the proposed mechanism in \ISS.  This
would contradict the fact that the low energy theory has $N-2$
supersymmetric vacua.

	Gluino condensation in the low energy $SO(N)$ Yang-Mills
theory generates the superpotential
\eqn\Wlo{W_{\rm lo} = \half (N-2) \Bigl[ 16 \Lambda_{\rm lo}^{3(N-2)} 
\Bigr]^{1/(N-2)}, }
where the low energy scale is fixed by the matching relation
$\Lambda_{\rm lo}^{3(N-2)} = m^{N+2} \Lambda^{2(N-4)}$.  This result
can be recovered from the effective superpotential
\eqn\Wupstairs{W_{\rm conf} = - \left[ {O_2 ^{N+2}\over
(N+2)^{N+2}\Lambda ^{2N-8}} \right]^{1/4}} 
in the upstairs theory after adding $\Wtree=\half mO_2$ and
integrating out $O_2$.  This superpotential is of the form
\Wdynamical.  Since we have already argued that it is absent on the
Higgs branch, we conclude that there must exist another ``confining
branch'' of the theory on which $W_{\rm conf}$ is generated when
$m\neq 0$.
\foot{When other source terms besides $g_2 = \half m$ in the tree level 
superpotential in \addw\ are nonzero, more general confining phase
superpotentials involving $O_{n>2}$ must be generated.  The confining
branch must also include the vacua with different unbroken $SO(m_i)$,
including the ones with $m_i=2$ which have a massless photon.}

	The multi-branch structure of the $\SON$ model's moduli space
is reminiscent of that for $SU(2)$ with two adjoints \refs{\ISthree,
\IStwo}.  We sketch its basic features in fig.~1.  The plane in the
figure represents the Higgs branch which is strongly coupled inside
the shaded region nearby the origin and weakly coupled in the domain
far away from $\vev{O_n} = 0$.  Partial free electric subspaces within
the Higgs branch which intersect the origin are illustrated by the
diagonal line in fig.~1.  When $m \neq 0$, the Higgs branch is lifted,
and the theory resides on the confining branch represented by the cone.
For any fixed value of $m$, the vacuum on the confining
branch consists of $N-2$ points with a mass gap.  As a result, there
are no massless moduli associated with the confining branch.  As can
be seen from the expectation value $\vev{O_2} = (N+2) (16 m^4
\Lambda^{2N-8})^{1/(N-2)}$, small values for $m$ yield vacua which lie
within the strongly coupled region nearby the moduli space origin.
The spreading of the cone with increasing $m$ mimics the behavior of
$\vev{O_2}$.  Points located on the confining branch at $m \gg
\Lambda$ are still strongly coupled, for 
the low energy limit of the $\SON$ 
model at such points is a super Yang-Mills theory.  The entire cone is
thus shaded grey.

\topinsert
\vskip -1cm
\epsfysize=13truecm \epsfbox[90 235 600 560]{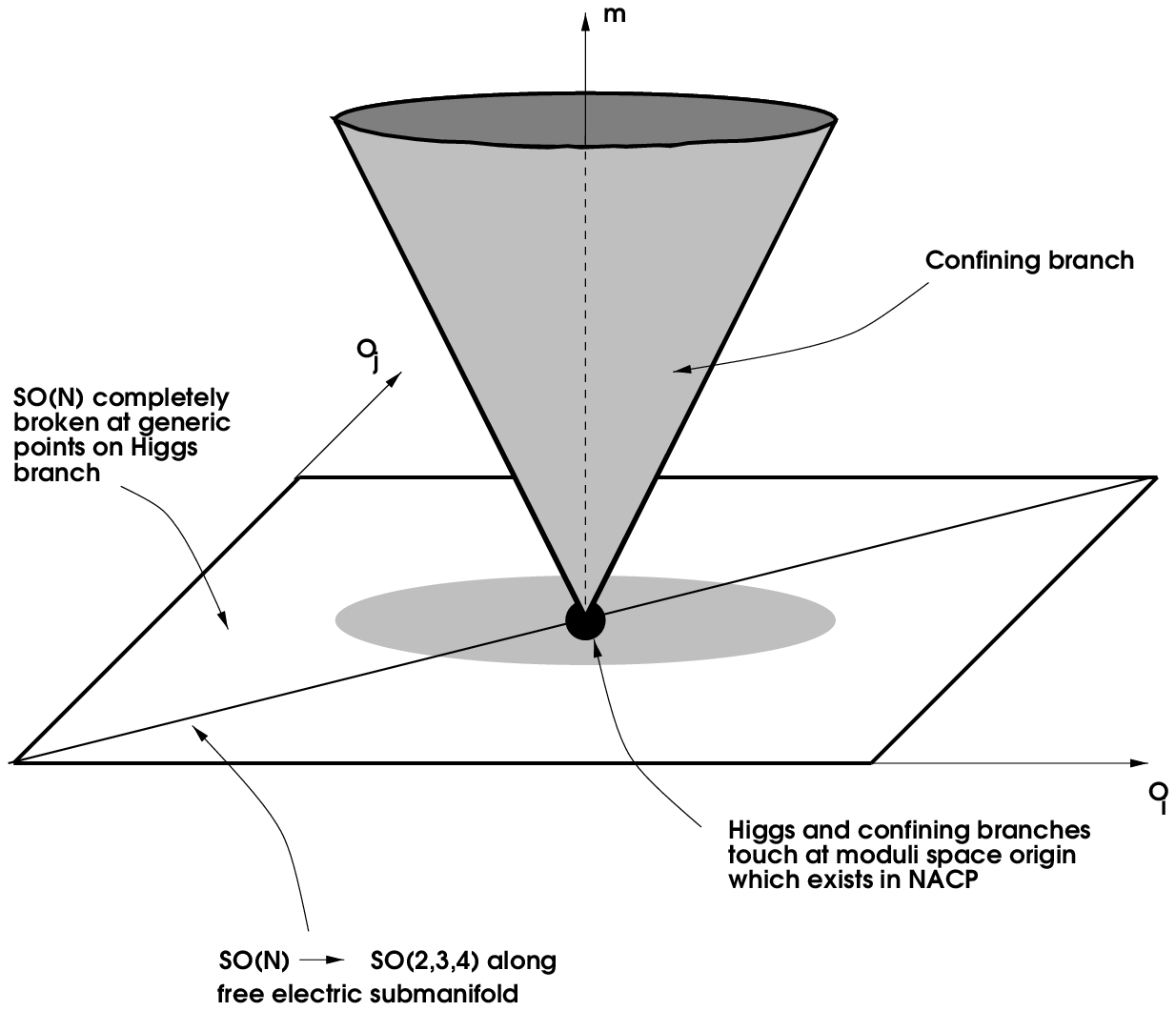}
  \hbox{\hskip 0.5truecm}
\centerline{Fig. 1: A schematic picture of the $\SON$ model's quantum 
moduli space.}
\bigskip
\endinsert

	In fig.~1, the Higgs and confining branches are shown touching
at the origin much as in $SU(2)$ theory with two adjoints
\refs{\ISthree, \IStwo}.  
The origin must then reside in an interacting 
non-Abelian Coulomb phase, for no free field spectrum could
incorporate both the Higgs and confining branches at this point.

	It is also possible, though unlikely, that the distinct Higgs
and confining branches for the $\SON$ model do {\it not} meet.  Such a
disconnected branch structure actually does occur in $SO(\Nc)$ theory
with $\Nf=\Nc-4$ vector flavors \ISthree.  Two inequivalent branches
arise from aligned and misaligned gaugino condensates within the low
energy unbroken $SO(4) \cong SU(2)\times SU(2)$ gauge group.  On the
Higgs branch, $\Wdyn$ vanishes as a result of cancellation between the
two gaugino condensates, 't~Hooft anomalies match at the origin and
the low energy spectrum contains only the massless moduli.  On the
confining branch, a destabilizing superpotential is dynamically
generated.  When $\Wtree \neq 0$, the Higgs branch ceases to yield
supersymmetric vacua and is eliminated, and the low energy theory
lives on the confining branch.  Since gaugino condensates do not
appear at generic points within our $\SON$ model's moduli space where
the gauge group is completely broken, we do not expect the theory with
a symmetric matter field to exhibit such a disconnected branch
structure.

	More generally, we expect that the theories recently classified in
\Manohar\ for which the gauge group can (can not) be completely broken 
do not (do) possess disconnected Higgs and confining branches
associated with aligned and misaligned gaugino condensates in the
unbroken product gauge group.  The general distinction between the two
cases may be seen by turning on a mass term $m$.  On the confining
branch, the expectation value of a matter field $Q$ with index $\mu$
is fixed by symmetries and holomorphy to be $\vev{Q} \sim (m^{\mu-G}
\Lambda^{3G-\mu})^{1/2G}$, where $G$ denotes the adjoint index.  If
$\mu < G$, the gauge group generally breaks to a non-trivial subgroup
and the moduli run off to infinity as $m \to 0$.  The confining branch
can then be thought of as a cone turned upside-down relative to that
in fig.~1.  The cone approaches, but never touches, the plane at
infinite moduli values.  The Higgs and confining branches are
therefore disconnected when $\mu < G$.  In contrast, it seems more
likely that the Higgs and confining branches intersect at the origin
in a non-Abelian Coulomb phase for $\mu > G$ theories where the gauge
group can be completely broken.  The $\SON$ model with a single
symmetric matter field falls into the latter category.

	The structure of the theory at the origin can be 
probed by perturbing the $\SON$ model with $\Wtree = \lambda \Tr \,
S^{k+1}$ for $k \ge 2$.  
The equations of motion for the eigenvalues of $S$ are $\lambda
(k+1) z^k+g_1=0$, where $g_1$ is a Lagrange multiplier that enforces the
tracelessness condition.  These equations generally only have the
trivial solution $z=g_1=0$.  The superpotential thus lifts the moduli
space.  When $N=mk$ with $m$ an integer, there is a one complex
dimensional moduli space of vacua which looks like
$\ev{S}=z\rm{diag}(e^{2\pi i \ell/k})\otimes 1_{m\times m}$ where $z
\in \BoldC$ and $\ell=1, \cdots, k$.  This vev breaks $\SON \to
SO(m)^k$, and all massless matter fields are eaten.  For all values of
$N$ and $k$, the theories with $\Wtree = \lambda \Tr S^{k+1}$ can be
regarded as the $N_f=0$ limit of a class of $\SON$ models analyzed in
ref.~\Intriligator\ with a single symmetric matter field and $\Nf$
vectors.  It has a dual description in terms of an $SO(4k-N)$ gauge
group, a traceless symmetric tensor $\widetilde S$ and a tree level
superpotential $\Wdualtree = {\tilde \lambda} \Tr\, {\widetilde
S}^{k+1}$.  This duality follows from the tower of dual pairs found in
\Intriligator\ for general $\Nf$ after all vector matter fields are
given masses and integrated out. 

	A special case occurs when $N=2k$.  $\Wtree$ then respects the
anomaly free $U(1)_R$ symmetry, and the electric and magnetic theories
are identical.  A one complex dimensional moduli space remains
unlifted by $\Wtree$ along which the gauge group is broken to
$SO(2)^k$.  The massless spectrum consequently contains $k$ photons.
In the particular case of $SO(4)\cong SU(2)\times SU(2)$ theory with
$S \sim ({\bf 3,3})$ and $\Wtree=\lambda S^3$, a two-dimensional
manifold of non-trivial $\CN=1$ renormalization group fixed points
emanates out from the origin as a function of $g_1$, $g_2$, and
$\lambda$ \Leigh.  For $N=2k>4$, there can not be an analogous
non-trivial RG fixed point at the origin of the moduli space since the
theories are asymptotically free with a dynamical scale $\Lambda$.
However, there could be non-trivial RG fixed points for $N=2k>4$ away
from the origin, with a line of fixed points as a function of
$\lambda$ and $\ev{S}/\Lambda$.  This scenario requires the
spontaneous breaking of scale invariance for $\ev{S}\neq 0$ and the
explicit breaking due to $\Lambda$ to cancel.  The duality of
\Intriligator\ would then provide a dual description of this fixed
point.

	For general $N$ and $k$, $\Wtree$ ($\Wdualtree$) breaks
$U(1)_\R$ in the electric (magnetic) theories.  In the far infrared,
an accidental $U(1)_R$ symmetry must be restored as part of the
super-conformal algebra of the (possibly free) IR fixed point.  When
the gauge coupling is strong, the appropriate $U(1)_R$ symmetry in the
same super-multiplet as the stress tensor should be close to the
anomaly free one.  The statement that the superpotential generally
violates this $U(1)_R$ symmetry is then equivalent to saying that it
is generally relevant or irrelevant, rather than marginal.  In cases
where the dual $SO(4k-N)$ gauge group is not asymptotically free and
the magnetic superpotential also appears irrelevant $(D(\Wdualtree) =
\threehalves R(\Wdualtree) \Longrightarrow 2k \le N)$, it seems
likely that the dual becomes free in the infrared.  The requirement
that these dual or free magnetic theories be recovered when the
original $\SON$ model is perturbed again favors the hypothesis that
various moduli space branches meet at the origin in a non-Abelian
Coulomb phase, which allows for nontrivial low energy dynamics.

	To recapitulate, we have given three arguments for why the
$\SON$ model's moduli space origin exists in an interacting
non-Abelian Coulomb phase.  Firstly, it is difficult to reconcile the
't~Hooft anomaly matching results with the existence of free electric
fields along subspaces that intersect the origin.  Secondly, the
moduli space must have a confining branch when the symmetric matter
field is given a nonzero mass.  This confining branch most likely
touches the Higgs branch at the origin.  Finally, a nontrivial phase
and branch structure must arise when the original $\SON$ model is
perturbed with a general tree level superpotential.  Perhaps a dual
description of the RG fixed point at the origin can be found which
would provide a weak coupling picture for the confining branch and
other phenomena associated with $\Wtree \neq 0$.  At present, no such
dual is known.

	In light of all these findings, we conclude that the $\SON$
model represents a rare example where global anomaly matching does not
signal simple confinement.  The existence of an infinite chain of
theories where anomaly matching appears to be misleading represents a
worthwhile point to bear in mind when analyzing the infrared behavior
of other supersymmetric and nonsupersymmetric models.

	If the $\SON$ theory does not confine, why do the anomalies
match?  Some insight into this coincidence can be gained from the
negative dimensional group theory relationship
\eqn\sonspnduality{SO(2N) = \bar{Sp(-2N)}}
where the overbar indicates interchange of symmetrization and
antisymmetrization \refs{\Parisi{--}\Dunne}.  This formal expression
links orthogonal groups acting on conventional bosonic tensor spaces
to negative dimensional symplectic groups defined by their action on
tensors in Grassmann vector spaces.  For instance, the dimension of an
$SO(2N)$ irrep labeled by a Young tableau $\lambda$ can be obtained up
to a sign from that for the corresponding $Sp(2N)$ irrep with the
transposed tableau $\lambda^\T$ by simply setting $N \to -N$.
Similarly, every $SO(2N)$ invariant scalar is related to an $Sp(2N)$
counterpart by replacing the symmetric $g_{\a\b}$ metric with its
antisymmetric $J_{\a\b}$ analogue and swapping $N \to -N$.

	The group theory relation in \sonspnduality\ suggests that the
$\SON$ model is mathematically similar to the theory with symmetry
group
\eqna\spnmodel
$$ \eqalignno{\Gdual & = \SpN_{\rm local} \times U(1)_\R & \spnmodel
a}$$
and matter content
$$ \eqalignno{A & \sim \Bigl( \> \anti \> ; -{4 \over 2N-2} \Bigr)
		& \spnmodel b} $$
which was studied in refs.~\refs{\ChoKraus,\CSS}.  Like its orthogonal
counterpart, this symplectic supersymmetric model possesses $N-1$
complex flat directions which are labeled by the operators
\eqn\Odualops{{\widetilde O}_n = \Tr (AJ)^n, \qquad n=2,3,\cdots, N.}
At generic points in moduli space, the expectation value for $A$
breaks $\SpN \to SU(2)^N$.  The quantum moduli space consequently has
a variety of disconnected branches which are associated with the
different possible signs for the gaugino condensates in each of the
$SU(2)$ factors of the unbroken gauge group.  The condensate sum
generally yields non-vanishing dynamical superpotentials, which lift
the classical vacuum degeneracy.  But on one branch, the different
condensate contributions precisely cancel and $\Wdyn=0$.
\foot{After taking into account scale 
matching relations, one can check that there is a Higgs branch with
such a cancellation for all $N$.}
The moduli space on this Higgs branch is smooth in terms of the
${\widetilde O}_n$ fields, and the `t~Hooft anomalies
\eqn\spnanomalies{\eqalign{
A_{U(1)_\R} &= -2 N-1 \cr
A_{U(1)^3_\R} &=  {(-2N-1)(5 N^2 + 2 N+ 1) \over (-N+1)^2} \cr}}
match at the microscopic and macroscopic levels.  Comparing these
anomaly expressions with their analogues in \hadronanomalies, we
observe
\eqn\anomrelns{\eqalign{
A_{U(1)_\R}^{Sp(2N)} & \qquad {\buildrel 2N \to -N \over
\longrightarrow}
\qquad A_{U(1)_\R}^{SO(N)} \cr
A_{U(1)^3_\R}^{Sp(2N)} & \qquad{\buildrel 2N \to -N \over 
\longrightarrow} \qquad A_{U(1)^3_\R}^{SO(N)}. \cr}}
So matching of the $U(1)_\R$ and $U(1)_\R^3$ anomalies in the $\SON$
theory with a single symmetric matter field appears to be an automatic
mathematical consequence of the same anomaly matchings in the $\SpN$
theory with an antisymmetric field.  In the symplectic model, the
matching is physically significant and signals genuine confinement.

	Given the connection between the $SO$ and $Sp$ anomalies, it
is amusing to note that $Sp$ analog of the discrete anomaly matching
in the $\SON$ model does {\it not} generally work.  In the $\SpN$
theory, a $Z_{2N-2}$ subgroup of the classical $U(1)_\A$ group, which
assigns the $A$ field charge 1, remains intact at the quantum level.
The $Z_{2N-2}^3$, $Z_{2N-2}^2 U(1)_\R$ and $Z_{2N-2} U(1)_\R^2$
anomalies, which need not match since they can be corrupted by the
massive spectrum, indeed do not match.  Furthermore, the difference
between the microscopic and macroscopic values for the
$Z_{2N-2}$-gravity-gravity anomalies
\eqn\discretespnanomalies{\eqalign{
A^{(\rm parton)}_{Z_{2N-2}} &= (2N^2-N-1)(1) = (N-1)(2N+1) \cr
A^{(\rm hadron)}_{Z_{2N-2}} &= \sum_{n=2}^N n = \half(N-1)(N+2) \cr}}
is $\Delta = {3 \over 2} N(N-1)$.  So the $Z_{2N-2}$ anomalies match
modulo $N-1$ if $N$ is even, but they fail to match if $N$ is odd.
	
	Instances where naive discrete anomaly matching arguments fail
have previously been noted \CsakiMurayama.  For example, the
nonvanishing vev for the baryonic glueball operator $B=\e_{\mu_1
\cdots \mu_{\Nc}} W^{\mu_1
\mu_2} W^{\mu_3 \mu_4} V^{\mu_5} \cdots V^{\mu_{\Nc}}$ within $SO(\Nc)$
theory with $\Nf = \Nc-4$ vectors leaves intact all continuous global
symmetries but breaks the instanton induced $Z_{2\Nc-8}$ down to
$Z_{\Nc-4}$.  Anomalies therefore match only for the latter discrete
subgroup and not for its larger progenitor.  A similar phenomenon may
resolve the discrete anomaly mismatch in the symplectic model.
\foot{We thank Csaba Cs\`aki for discussions on this point.}
The $N$ glueballs which must emerge when $\SpN \to SU(2)^N$ come from
linear combinations of $B_n = \Tr (WJ)^2 (AJ)^n$ where $0 \le n
\le N-1$.  The only one of these operators whose vev does not break
$U(1)_\R$ is $B_{N-1}$.  If this glueball composite develops a nonzero
expectation value, all anomalies involving the unbroken discrete group
$Z_{N-1}$ match between the microscopic and macroscopic theories for
$N$ even.  Moreover, anomalies also match for $N$ odd provided the low
energy sigma model contains an odd number of massive Majorana fermions
with charge $N-1$.  It is important to note that only an even number
of such Majorana fermions can exist when $N$ is even if the prior
anomaly results are not to be disrupted.  This rather involved
scenario appears to yield viable anomaly matching results within the 
symplectic model.

\bigskip\bigskip\bigskip\bigskip
\centerline{{ \bf Acknowledgments}}
\bigskip

JB thanks Lance Dixon and Lisa Randall for discussions.  PC thanks Per
Kraus for collaborating at an early stage on this work and Philip
Argyres, Howard Georgi, Martin Schmaltz, Nathan Seiberg, Matt
Strassler and Sandip Trivedi for sharing their insights.  KI thanks
Nathan Seiberg for useful discussions and Gustavo Dotti and Aneesh
Manohar for rekindling his interest in these models.  We also
acknowledge support from the Dept. of Energy under Grant
DOE-FG02-91ER40671 (JB), the National Science Foundation under Grant
\#PHY-9218167 (PC), and the Dept. of Energy under Grant
DOE-FG03-97ER40506 and the Alfred Sloan Fellowship Foundation (KI).

\listrefs
\bye